\documentclass[aps,pra,twocolumn,showpacs,superscriptaddress,floats]{revtex4}

\usepackage[dvips,final]{graphics}
\usepackage{amssymb}

\begin{document}

\newcommand{\no}{\nonumber}
\newcommand{\etal}{{\em et~al }}
\newcommand{\ie}{{\it i.e.\/}\ }

\title{\center{Pulsed squeezed vacuum characterization without
homodyning}}

\author{J\'er\^ome Wenger}
\affiliation{Laboratoire Charles Fabry de l'Institut
d'Optique, CNRS UMR 8501, F-91403 Orsay, France}

\author{Jarom\'ir Fiur\'a\v{s}ek}
\affiliation{QUIC, Ecole Polytechnique, CP 165, Universit\'e libre
de Bruxelles, 1050 Bruxelles, Belgium} \affiliation{Department of
Optics, Palack\'y University, 17. listopadu 50, 77200 Olomouc,
Czech Republic}

\author{Rosa Tualle-Brouri}
\affiliation{Laboratoire Charles Fabry de l'Institut
d'Optique, CNRS UMR 8501, F-91403 Orsay, France}

\author{Nicolas J. Cerf}
\affiliation{QUIC, Ecole Polytechnique, CP 165,
Universit\'e libre de Bruxelles, 1050 Bruxelles, Belgium}

\author{Philippe Grangier}
\affiliation{Laboratoire Charles Fabry de l'Institut
d'Optique, CNRS UMR 8501, F-91403 Orsay, France}

\begin{abstract}
Direct photon detection is experimentally implemented to measure
the squeezing and purity of a single-mode squeezed vacuum state
without an interferometric homodyne detection. Following a recent
theoretical proposal [arXiv quant-ph/0311119], the setup only
requires a tunable beamsplitter and a single-photon detector to
fully characterize the generated Gaussian states. The experimental
implementation of this procedure is discussed and compared with
other reference methods.
\end{abstract}
\maketitle

\section{Introduction}

Squeezed states of light play an important role in the development
of quantum information processing with continuous variables
\cite{Braunsteinbook}, where the information is encoded in
two conjugate quadratures of an optical field mode. These states
may for example be used as a main resource for
quantum cryptographic protocols (see \cite{pra,prl} and references
therein). They may also serve as an entanglement source
since combining two squeezed states at a beamsplitter creates an
entangled two-mode squeezed state such as those required for quantum
teleportation \cite{teleport} or dense coding \cite{densecod}.
In addition, squeezing has been shown to be an irreducible
resource for realizing an arbitrary linear canonical transformation
\cite{braunstein}.

\par
Any attempt to process squeezed states in quantum communication or
computation systems will necessarily face the problem of
characterizing these states. A possible complete description of a
general quantum state is obtained by reconstructing its Wigner
function using quantum tomographic procedures
\cite{ulf,Breit97,Welsch99}. Alternatively, for a Gaussian state,
which is  fully described by its first and second order
moments, another complete characterization is provided by the mean
values of the conjugate quadratures $x$ and $p$ together with the
associated covariance matrix $\gamma$. From this, one may compute
various relevant parameters such as the maximum observable
squeezing \cite{Simon94} or the degree of purity \cite{Kim02,Paris03}.

\par

In this paper, we follow an idea originally due to two of the present
authors that consists of measuring the squeezing
and purity of a Gaussian state without homodyne detection, that is,
without any strong local oscillator beam providing a phase reference
\cite{JFNJCquantph}.
The suggested setup only relies on beamsplitters and single-photon
detectors. It generally requires a joint measurement of two copies of
the Gaussian state, but single-copy measurements suffice if it is
{\it a priori} known that the mean values of the quadratures vanish.
Thus, in the latter case,
which actually applies to all quantum information schemes based on
squeezed vacuum states, no interferometric stability is required
to determine the squeezing and purity, unlike with homodyne detection
schemes.
\par

Hereafter, we will focus on the important case of a single-mode
squeezed vacuum state and discuss the experimental feasibility and
relevance of this photon-counting characterization procedure. Some
useful notations to describe a squeezed vacuum beam are introduced
in section~\ref{S2}. Section~\ref{S2bis} then presents the
experimental setup together with two classical and homodyne
measurement procedures that are used as a reference to
characterize the generated squeezed vacuum states. In
Section~\ref{S1}, we briefly review the photon-counting
characterization method applied to the special case of a
single-mode Gaussian state (more details can be found in
\cite{JFNJCquantph}). In Section~\ref{S3}, we present the
experimental results of this characterization method, while
Section~\ref{S4} discusses the constraints on the global detection
efficiency that are put by this method. Numerical simulations are
used to illustrate the photon-counting method for values of the
global detection efficiency that are presently unreachable in the
experiments.

\section{Squeezed vacuum description}
\label{S2}

Theoretically, a general Gaussian state with zero mean values of
quadratures is fully characterized by its covariance matrix
$\gamma$, which comprises the second moments of the conjugate
quadratures $x=a+a^\dagger$ and $p=(a-a^\dagger)/i$, with
$[x,p]=2i$. For states with zero mean values of quadratures, the
covariance matrix $\gamma$ can be expressed as follows,
\begin{equation}
\gamma=\left(
\begin{array}{cc}
\langle x^2\rangle & \frac{1}{2}\langle x p + p x\rangle
\\[2mm]
\frac{1}{2}\langle x p + p x \rangle & \langle  p^2 \rangle
\end{array}
\right).
\label{gammadefinition}
\end{equation}
In order to  determine the squeezing and purity, we only need to
measure the two invariants of the covariance matrix, namely the
trace $\mathrm{Tr}(\gamma)$ and determinant $\det(\gamma)$.

\par
\begin{figure}
\center
\includegraphics{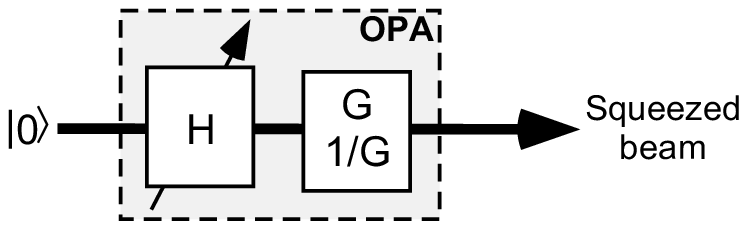}
\caption{General single-mode Gaussian state generation. $H$
denotes the intensity gain of the phase-independent amplifier and
$G$ stands for the intensity gain of the phase-dependent
amplifier.} \label{expmodel}
\end{figure}

From a more physical point of view, one can use the fact that the
most general single-mode Gaussian state with $\langle
x\rangle=\langle p\rangle=0$ can be expressed as a squeezing
operator applied to a Gaussian thermal state
\cite{Adam95,Paris03}. Translated into an optical setup, this is
implemented by two simple linear amplifiers as depicted in
Fig.~\ref{expmodel}~: a \textit{phase-insensitive} amplifier of
gain $H$ followed by a \textit{phase-sensitive} amplifier of gains
$G$ and $1/G$ (in the following we take $G, H >1$). In other
words, the physics of the optical parametric amplifier (OPA) can
be modeled by a ``black box'' squeezer which is parametrized by
$H$ and $G$. These two parameters are equivalent to the two
phase-insensitive parameters $\mathrm{Tr}(\gamma)$ and
$\det(\gamma)$ of the Gaussian state generated by the black box
from the vacuum.

Let us describe the transformation effected by the amplifier
depicted in Fig.~\ref{expmodel}. First, one can express the
conjugate quadrature variables at the output of the OPA as
\begin{eqnarray}
  x_{\mathrm{out}} & = & \frac{1}{\sqrt{G}} ( \sqrt{H} x_{\mathrm{vac}}
  + \sqrt{H-1} x_{\mathrm{anc}}), \\
  p_{\mathrm{out}} & = & \sqrt{G} ( \sqrt{H} p_{\mathrm{vac}}
  - \sqrt{H-1} p_{\mathrm{anc}}),
\end{eqnarray}
where we chose $x_{\mathrm{out}}$ ($p_{\mathrm{out}}$) as the
squeezed (anti-squeezed) quadrature, and $x_{\mathrm{vac}}$ and
$x_{\mathrm{anc}}$ denote the vacuum and ancilla quadratures at
the input ports of the total amplifier, respectively. One can then
express the variances of the squeezed and anti-squeezed
quadratures at the output of the OPA as
\begin{eqnarray}
  V_{\mathrm{min}} & = & (2 H-1)/G, \\
  V_{\mathrm{max}} & = & (2 H-1)\; G.
\end{eqnarray}
The trace and determinant of the covariance matrix read
\begin{eqnarray}
 \mathrm{Tr}(\gamma)&=&V_{\mathrm{min}}+V_{\mathrm{max}}, \\
 \det(\gamma)&=&V_{\mathrm{min}}V_{\mathrm{max}}.
\end{eqnarray}
This system of
equations can be inverted, and the squeezed and anti-squeezed variances
can be expressed
in terms of the trace and the determinant of $\gamma$,
\begin{equation}
V_{\mathrm{max,min}}=\frac{1}{2}
\left[\mathrm{\mathrm{Tr}}(\gamma)\pm\sqrt{\mathrm{\mathrm{Tr}}^2(\gamma)-4\det(\gamma)}\right].
\label{lambdasinglemode}
\end{equation}
Finally, the purity $\mathcal{P}=\mathrm{\mathrm{Tr}}[\rho^2]$ of
a mixed state $\rho$ is, for any single-mode Gaussian state,
directly linked to the average photon number of thermal noise
$\overline{n}=H-1$~:
\begin{equation}
   \mathcal{P} = \frac{1}{2\overline{n}+1} =\frac{1}{2H-1}.
\end{equation}
Equivalently, in terms of the covariance matrix, we have
\begin{equation}
\mathcal{P}=\frac{1}{\sqrt{\det(\gamma)}}.
\label{puritysinglemode}
\end{equation}
\par

\section{Reference classical and homodyne characterization methods}
\label{S2bis}

\begin{figure}
\center
\includegraphics{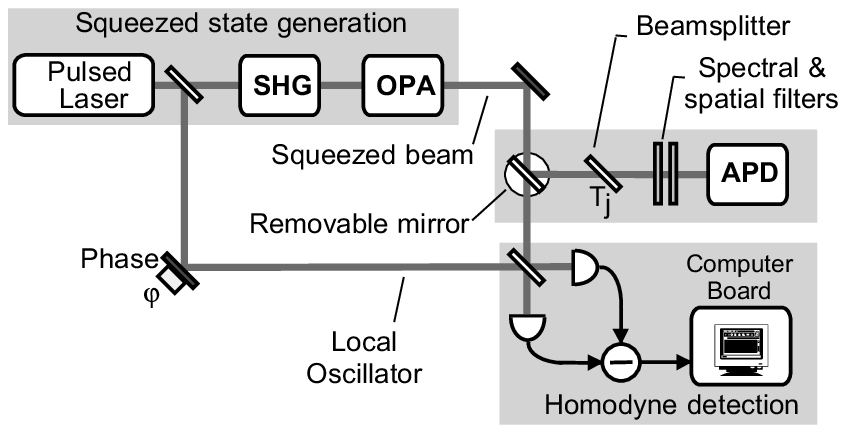}
\caption{Simplified experimental setup. SHG : second harmonic
generation module; OPA : degenerate optical parametric amplifier;
APD : avalanche photodiode photon counting module.}
\label{expsetup}
\end{figure}

A new scheme for pulsed squeezed light generation has recently
been developed \cite{WengerOL} and will be used here to compare
the photon-counting characterization method to standard methods.
The experimental setup is depicted in Fig.~\ref{expsetup}. The
initial pulses are obtained from a titanium-sapphire laser
delivering nearly Fourier-transform limited pulses centered on 846
nm, with a duration of 150 fs, a typical energy of 40 nJ, and a
repetition rate of 780.4 kHz. These pulses are frequency doubled
in a single pass through a thin (100 $\mu$m) crystal of potassium
niobate (KNbO$_3$), cut and temperature-tuned for non-critical
type-I phase-matching. The second harmonic power is large enough
to obtain a significant single-pass parametric gain in a similar
KNbO$_3$ crystal used in a type-I spatially degenerate
configuration.
\par

The squeezed beam can then be directed onto two different
detection modules using a removable mirror:
\par
{\it (i) Homodyne detection module~:} the squeezed vacuum beam
interferes with the local oscillator beam (LO) in a balanced
homodyne detection setup. A main feature of our experiment is that
all the processing is done in the time domain, not in the
frequency domain. For each incoming pulse, the fast acquisition
board samples one value of the signal quadrature in phase with the
local oscillator \cite{WengerOL}.
\par
{\it (ii) Photon counting module~:} the squeezed vacuum beam is
transmitted via a beam splitter of tunable transmittance $T$ and
then passes through a spatial filter (made of two
Fourier-conjugated pinholes) and a 3 nm spectral filter centered
at the laser wavelength, before being detected by a silicon
avalanche photodiode (APD).
\par

\begin{figure}[!ht]
\center
\includegraphics{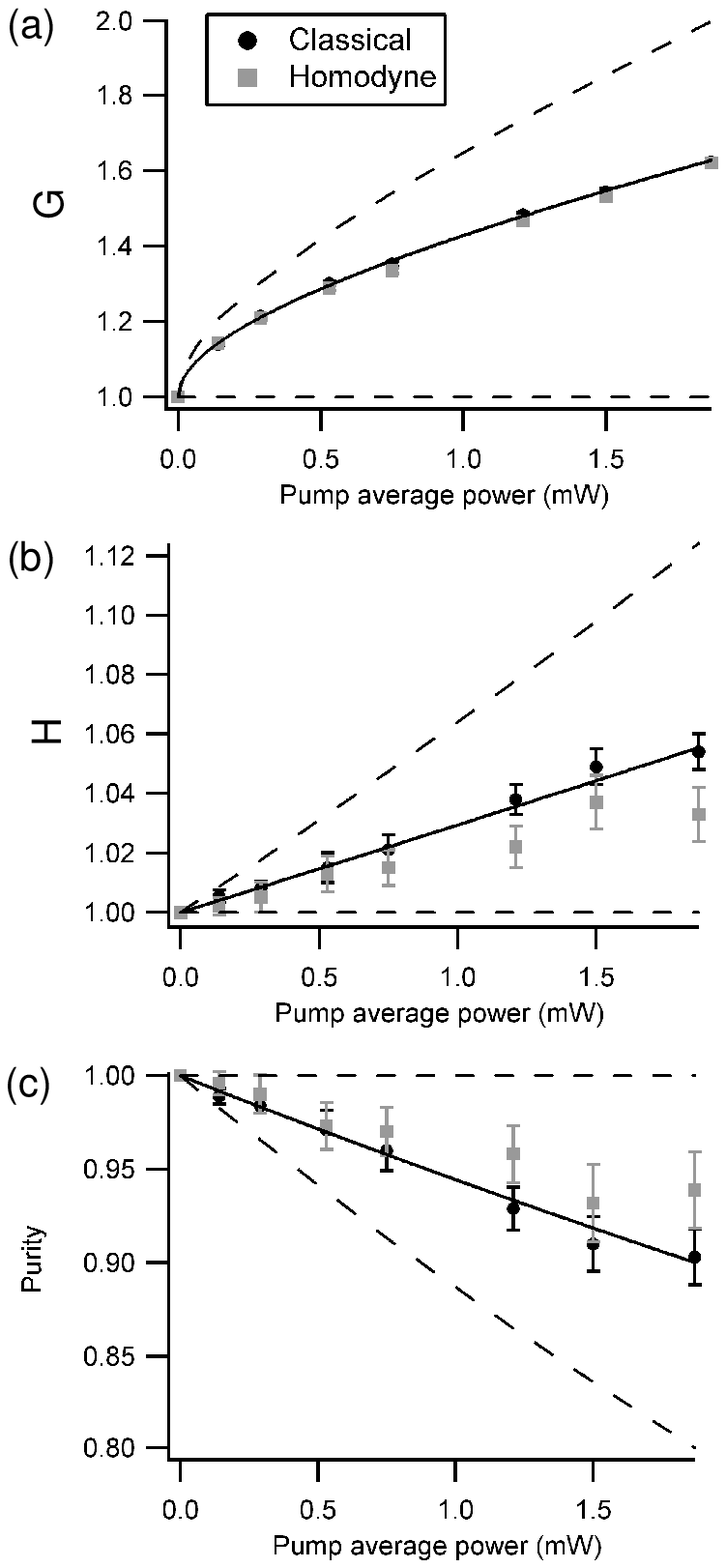}
\caption{Parametric gains $G$, $H$ and purity $\mathcal{P}$ versus
average pump power at 423nm. The solid line corresponds to a fit
on the classical results according to plane wave theory.
``Classical" stands for the classical probe gain measurements
(black disks), ``Homodyne" stands for the balanced homodyne
detection variance measurements (gray squares). The bounds
inferred from the photon counting method, to be described in
section \ref{S1}, are indicated by the two dashed lines (see
further explanations in section \ref{S3}).} \label{expresult1}
\end{figure}

\begin{figure}[!ht]
\center
\includegraphics{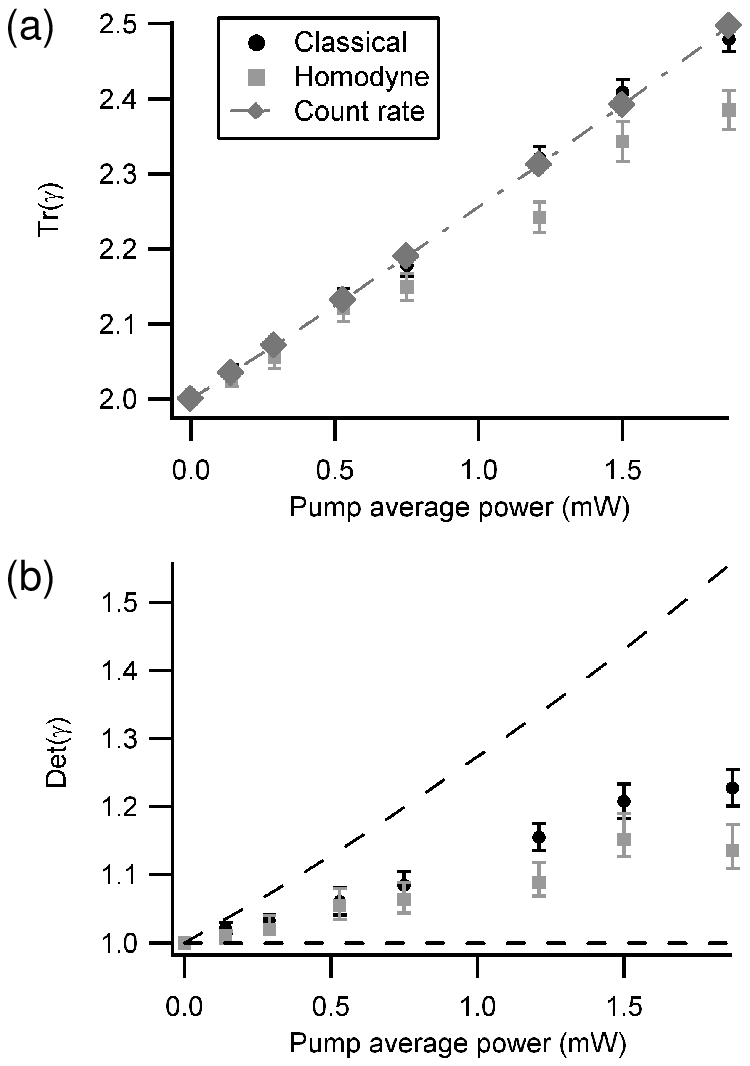}
\caption{Trace $\mathrm{Tr}(\gamma)$ and determinant
$\det(\gamma)$ versus average pump power at 423nm. The annotations
are the same as in fig.\ref{expresult1}. For ease of viewing, the
trace values obtained from the photon counting method (gray
diamonds) are linked by a dash-dotted line in Fig.(a). In Fig.(b),
the two dashed lines indicate the limits on the determinant
knowledge obtained from the photon counting method (see further
details in section \ref{S3}).} \label{expresult2}
\end{figure}

To start the characterization procedure, a basic measurement is to
monitor the classical amplification and de-amplification of a
probe taken from the fundamental beam. This is easily done by
direct detection of a probe beam averaged power on a photodiode.
Setting the relative phase between the probe and the second
harmonic pump beam allows to tune the classical gain from the
minimum de-amplification intensity gain
$\mathcal{G}_{\mathrm{min}}$ to the maximum gain
$\mathcal{G}_{\mathrm{max}}$. The measurement of the classical
gains $\mathcal{G}_{\mathrm{min}}$ and
$\mathcal{G}_{\mathrm{max}}$ gives an estimate of $G$ and $H$,
\begin{eqnarray}
  G & = & \sqrt{\mathcal{G}_{\mathrm{max}}/\mathcal{G}_{\mathrm{min}}}, \\
  H & = & \sqrt{\mathcal{G}_{\mathrm{max}}
  \mathcal{G}_{\mathrm{min}}}.
\end{eqnarray}
The experimental results of the squeezed vacuum characterization
for different values of the pump power are shown in
Figs.~\ref{expresult1} and \ref{expresult2}, marked as
``Classical'' (black disks).

Following the principle of quantum tomography, a powerful approach
is to completely characterize the squeezed vacuum by conjugate
quadratures homodyne measurements. The time-resolved balanced
homodyne detection allows to measure the squeezed and
anti-squeezed quadrature variances $V_{\mathrm{hom,min}}$ and
$V_{\mathrm{hom,max}}$. Imperfections and losses in this detection
are modeled by a beamsplitter of transmission
$\eta_{\mathrm{hom}}$ (in intensity). The procedure to measure the
detection efficiency is well established from squeezing
experiments \cite{laporta}, and it can be cross-checked by
comparing the classical parametric gain and the measured degree of
squeezing. We note the homodyne detection efficiency
$\eta_{\mathrm{hom}} = \eta_T \eta_H^2 \eta_D = 0.76 \pm 0.01$,
where the overall transmission $\eta_T = 0.92$, the mode-matching
visibility $\eta_H = 0.935$, and the detectors efficiency $\eta_D
= 0.945$ are independently measured. Given this efficiency, one
can correct for losses and deduce the squeezed and anti-squeezed
quadrature variances at the output port of the OPA, namely
\begin{eqnarray}
  V_{\mathrm{min}} & = & (V_{\mathrm{hom,min}}-1
  +\eta_{\mathrm{hom}})/\eta_{\mathrm{hom}}, \\
  V_{\mathrm{max}} & = & (V_{\mathrm{hom,max}}-1
  +\eta_{\mathrm{hom}})/\eta_{\mathrm{hom}}.
\end{eqnarray}
This allows the full characterization of the state parameters
($\mathrm{Tr}(\gamma)$, $\det(\gamma)$) or the OPA parameters
($G$, $H$) following the above formulae. The experimental results
of this second characterization method are also displayed in
Figs.~\ref{expresult1} and \ref{expresult2}, marked as
``Homodyne'' (gray squares). As one can notice in
Figs.~\ref{expresult1} and \ref{expresult2}, for high pump powers
the ``homodyne'' and ``classical'' values do not well overlap
within their respective error bars. A main reason for this is that
the ``black-box'' model developed above is basically a
\textit{single-mode} model, and thus suffers from fundamental
limitations, while for high pump powers the physics involved in
parametric de-amplification are known to fall into a multi-mode
regime \cite{Gid}.
\par

In the following, we will use these ``classical'' and ``homodyne''
methods as references to check the validity of the photon-counting
characterization.

\section{Photon-counting characterization method} \label{S1}

Let us briefly introduce the methods implemented here for
measuring the properties of a Gaussian state by photon counting,
which are derived from the original procedure presented in
\cite{JFNJCquantph}. We will restrict our attention to a
single-mode Gaussian state with zero coherent displacement,
$\langle x\rangle=\langle p\rangle=0$, which is the case in the
present experiment depicted in Fig.~\ref{expsetup}. The squeezed
vacuum mode impinges on a beam splitter with tunable transmittance
$T$ before being measured by an avalanche photodiode that is
sensitive to single photons and can respond with two measurement
outcomes, either a click or a no-click. This detector with overall
detection efficiency $\eta_{\mathrm{APD}}$ can be modeled as a
beam splitter with transmittance $\eta_{\mathrm{APD}}$ followed by
an ideal detector that performs a dichotomic measurement described
by the projectors $\Pi_0=|0\rangle\langle 0|$ (a no-click) and
$\Pi_1=\openone - \Pi_0$ (a click). In the rest of this section,
we assume that the detector is ideal, while $\eta_{\mathrm{APD}}
\ne 1$ can be taken into account by substituting $T \rightarrow
\eta_{\mathrm{APD}} T$.

The probability of no-click of an ideal detector is given by
$P=\langle 0|\rho|0\rangle$. It can be determined from the Husimi
Q-function which provides a phase-space representation of the state $\rho$:
$Q(\alpha)$ is defined as the overlap of $\rho$ with a coherent
state $|\alpha\rangle$.
The Q-function of a Gaussian state with zero mean values of quadratures
is a Gaussian function centered at the origin,
\begin{equation}
Q(r)=\frac{1}{ 2\pi \sqrt{\det (\gamma+I)}} \exp \left[
-\frac{1}{2} \; r^{T}(\gamma+I)^{-1}r \right], \label{Qfunction}
\end{equation}
where $r=(x,p)$ and $I$ is the identity matrix.
Since vacuum is just a special case of a coherent state, the probability of
projecting the state (\ref{Qfunction}) onto vacuum  reads  $P=4\pi \, Q(0)$
and, on inserting $r=0$ in Eq.~(\ref{Qfunction}), we obtain
\begin{equation}
P=\frac{2}{\sqrt{\det (\gamma+I)}}. \label{probability1}
\end{equation}

The characterization method works by carrying out measurements of
the probabilities of no-click $P_j$ for several different
transmittances $T_j$. The covariance matrix $\gamma_j$ of the
state after passing through the beam splitter reads
$\gamma_j=T_j\gamma+(1-T_j)I$, where $\gamma$ is the covariance
matrix of the input state.  On inserting $\gamma_j$ into
Eq.~(\ref{probability1}), we obtain after some simple algebra
\begin{equation}\label{eqPj}
\frac{4}{P_{j}^{2}}=T_j^2\, \det(\gamma) +T_j(2-T_j)\,
\mathrm{\mathrm{Tr}}(\gamma) + (2-T_j)^2.
\end{equation}

We thus find that $P_j$ depends on $T_j$ (or, more generally, on
$\eta_{\mathrm{APD}}T_j$) and  on the determinant  and trace of
the covariance matrix $\gamma$ of the input state. Note that
$4/P_j^2$ is a \emph{linear} function of the two unknown
quantities $\det(\gamma)$ and $\mathrm{Tr}(\gamma)$. Thus,
measurements of $P_j$ for only two different transmittances simply
suffice to determine the trace and the determinant, as the system
of linear Eqs. (\ref{eqPj}) can easily be solved and yields
\begin{equation}
\mathrm{\mathrm{Tr}}(\gamma)=\frac{2}{T_2-T_1}\left(\frac{T_2}{T_1P_1^2}
-\frac{T_1}{T_2P_2^2}\right)+2-\frac{2}{T_1}-\frac{2}{T_2},
\label{trgammasolution}
\end{equation}
\begin{equation}
\det(\gamma)=\frac{2}{T_1-T_2}\left(\frac{2-T_2}{T_1P_1^2}-\frac{2-T_1}{T_2P_2^2}
\right)+\frac{(2-T_1)(2-T_2)}{T_1T_2}. \label{detgammasolution}
\end{equation}
Then, having obtained the determinant and trace of $\gamma$, we
can determine the squeezing properties of the state from Eq.
(\ref{lambdasinglemode}) as well as its purity
(\ref{puritysinglemode}).

Dealing with a real world experiment, with unavoidable noises and
uncertainties, a more realistic procedure would consist in
performing the experiment for as many transmission values $T_j$ as
possible and then trying to get the most information from these
various measurements. One possibility to gain information from
more than two measurements is to implement a maximum-likelihood
(ML) parameter estimation method (for a review, see for instance
Refs. \cite{Hradil97,Paris00,Rehacek01}). This procedure provides
the values of the parameters $\mathrm{Tr}(\gamma)$ and
$\det(\gamma)$ that are the most likely to yield the observed
experimental data. In mathematical terms, this boils down to
finding the maximum of the joint probability density
\begin{equation}
\mathcal{L}\left(\mathrm{Tr}(\gamma),\det(\gamma)\right)
=\prod_{j=1}^n P_j^{N_{\mathrm{rep}}-C_j}(1-P_j)^{C_j}, \label{L}
\end{equation}
which is called the likelihood function of the given experimental
data. Here, $C_j$ denotes the number of photodetector clicks per
second for transmittance $T_j$ and $N_{\mathrm{rep}}$ is the pulse
repetition rate. The probability $P_j$ is linked to
$\mathrm{Tr}(\gamma), \det(\gamma)$ and $T_j$ by Eq.~(\ref{eqPj}).
Actually, we also have to take into account some additional
constraints on the parameters $\mathrm{Tr}(\gamma)$ and
$\det(\gamma)$. The fact that the covariance matrix $\gamma$ is
positive definite and must satisfy the generalized Heisenberg
uncertainty relation $\det(\gamma)\geq 1$ puts the constraints
\begin{equation} \label{Constraints}
    1 \leq \det(\gamma)\leq
    \left(\frac{\mathrm{Tr}(\gamma)}{2}\right)^2.
\end{equation}

Next section now presents and discusses the results of this
characterization procedure from its experimental implementation.

\section{Experimental results} \label{S3}

Hereafter, we denote by $T$ only the transmittance of the
(lossless) variable beamsplitter. Non-unit transmissions of the
spectral and spatial filters and imperfect detection efficiency
are taken into account by an overall efficiency parameter
$\eta_{\mathrm{APD}}$ of the APD detection.

\begin{figure}[!t!]
\center
\includegraphics{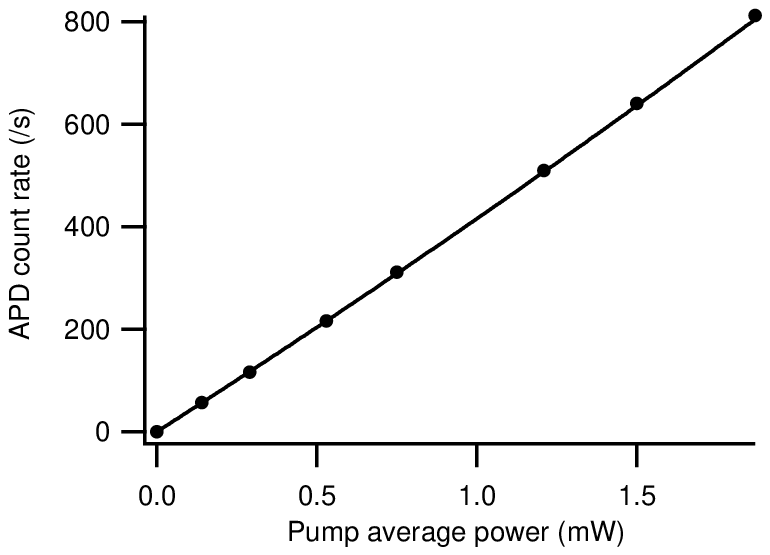}
\caption{Number of photon-detection events per second versus
average pump power for maximum transmission of the variable
beamsplitter $T=1$. The solid line is a fit following
Eq.(\ref{Nclicksexp}), from which we extracted an estimate of the
photon-couting detector efficiency $\eta_{\mathrm{APD}}=0.84\times
 10^{-2} \pm 0.013 \times 10^{-2}$. } \label{Nvspump}
\end{figure}

\par
A first step is to estimate this overall efficiency
$\eta_{\mathrm{APD}}$ in order to apply the characterization
method. Setting the supplementary beamsplitter to a transmittance
of 1, $\eta_{\mathrm{APD}}$ can be estimated from the measurement
of the number of photon-counting events detected per second
$N_{\mathrm{clicks}}$ for different pump powers. In the limit
of low $\eta_{\mathrm{APD}}$'s, the number of clicks
detected per second can be approximated as
\begin{eqnarray} \label{Nclicksexp}
  N_{\mathrm{clicks}} = {1\over 2} \eta_{\mathrm{APD}} N_{\mathrm{rep}}
  [ (H-1/2)(G+1/G)-1],
\end{eqnarray}
where the dependence of $G$ and $H$ versus the pump power
is obtained from the curve fit on the ``classical" results
presented Figs.~\ref{expresult1}(a) and (b), while
$N_{\mathrm{rep}}=780.4$kHz is the repetition rate. With our
experimental results (see Fig.~\ref{Nvspump}) and a repetition
rate of $780.4$ kHz, the fit of $N_{\mathrm{clicks}}$ versus the
pump power gives $\eta_{\mathrm{APD}}=0.84\times 10^{-2} \pm 0.013
\times 10^{-2}$. This value can be cross-checked with the overall
efficiency inferred from transmission factors of an intense probe
beam~: the spatial and spectral filters transmit respectively
$16$\% and $17$\% of the probe beam, while the APD quantum
efficiency is estimated to about $50$\%, leading to an overall
detection efficiency of the probe of about $1.4$\%. The difference
between the latter value and the above estimate of
$\eta_{\mathrm{APD}}$ may be explained by slight differences
between the modes of the probe (set for maximal classical
de-amplification) and the squeezed vacuum.
\par

In our experiment, we used between 4 and 6 different settings for
the beamsplitter transmittance $T_j$. For each $T_j$, we performed
100 measurements of the number of clicks per second to get a good
statistical accuracy on $C_j$. As a result of an appropriate
gating of the detection, the dark count rate remained reasonably
low (about 20$s^{-1}$) and was subtracted from the data.

As shown above, only two different settings of the beamsplitter
transmittance, $T_1$ and $T_2$, are enough to extract the value of
$\det(\gamma)$ and $\mathrm{Tr}(\gamma)$ following Eqs.
(\ref{trgammasolution}) and (\ref{detgammasolution}). Actually,
formula (\ref{trgammasolution}) indeed leads to an estimate of
$\mathrm{Tr}(\gamma)$ which is satisfactorily close to the values
obtained from either homodyne or classical measurement. However,
as far as the determination of $\det(\gamma)$ is concerned, the
formula (\ref{detgammasolution}) does not give any reliable
estimate. This results from the fact that, in the experiment, we
have to work with small detection efficiencies
$\eta_{\mathrm{APD}}\ll 1$ so that small uncertainties on $P_1,
P_2, T_1, T_2$ have much larger influence on $\det(\gamma)$ than
on $\mathrm{Tr}(\gamma)$. For instance, if we take the derivative
of $\mathrm{Tr}(\gamma)$ and $\det(\gamma)$ with respect to $P_1$,
we find,
\begin{eqnarray}
  \frac{d\; \mathrm{Tr}(\gamma)}{dP_1} & = &
  \frac{4}{\eta_{\mathrm{APD}}P_1^3}, \\
  \frac{d\det(\gamma)}{dP_1} & = &  \frac{-16}{\eta_{\mathrm{APD}}^2
  P_1^3}.
\end{eqnarray}
This shows that in our experimental setup, the determinant is
about 400 times more sensitive to small uncertainties on $P_1$
than the trace.

\par

In order to gain information on the determinant of the covariance
matrix as well as to increase the accuracy of the estimate of its
trace, we used the full set of measurements for the different
beamsplitter transmittances by performing a maximum-likelihood
estimation as introduced in the previous section. The logarithm of
the likelihood function $\mathcal{L}$ given by Eq.(\ref{L}) was
computed from the measured data, the above estimate of the overall
detection efficiency $\eta_{\mathrm{APD}}$, and the values of the
transmittance $T_j$ obtained from direct power transmission of the
probe beam. The global maximum of $\log(\mathcal{L})$ was then
found by brute force numerical search. The experimental results of
the estimated $\mathrm{Tr}(\gamma)$ for several different pump
powers are shown in Fig.~\ref{expresult2}(a), and fully coincide
with the values inferred from the classical gain measurements. Out
of the three trace-estimation procedures, the photon-counting
method associated with log-likelihood maximization provides the
lowest uncertainty on the result.
\par

Unfortunately, given the low detection efficiency
$\eta_{\mathrm{APD}}$ of our experimental setup, the likelihood
function is almost flat as a function of $\det(\gamma)$ in the
region that is allowed by the constraints~(\ref{Constraints}).
Consequently, no reliable estimate of the determinant could be
obtained from our experimental data, the log-likelihood
maximization method returning essentially a random value between 1
and $(\mathrm{Tr}(\gamma)/2)^2$. Similarly, our experimental data
only provides bounds on the parametric gains $G$ and $H$ given the
sole knowledge of the trace $\mathrm{Tr}(\gamma)$~:
\begin{eqnarray}
  1 & \leq  G  \leq & \left[ \frac{\mathrm{Tr}(\gamma)+\sqrt{\mathrm{Tr}(\gamma)^2-4}}{\mathrm{Tr}(\gamma)-\sqrt{\mathrm{Tr}(\gamma)^2-4}}
  \right]^{1/2}, \\
  1 & \leq  H  \leq & \frac{\mathrm{Tr}(\gamma)+2}{4}.
\end{eqnarray}
Thus, the dashed lines in Figs. \ref{expresult1}(a), (b), (c), and
\ref{expresult2}(b) take into account the fact that no
estimate of the determinant better than the bounds~(\ref{Constraints})
could be obtained by the photon-counting method given the
estimate of the trace. We also tried various other numerical
methods - such as least squares inversion - but neither provided a
reliable estimate of $\det(\gamma)$.
\par

Some better insight on the intrinsic difficulty to get an estimate
of $\det(\gamma)$ can be obtained by rewriting Eq.~(\ref{eqPj}) as
\begin{eqnarray}
  \frac{4}{P_j^2} & = &  \left( \det(\gamma)-\mathrm{Tr}(\gamma)+1 \right) \eta_{\mathrm{APD}}^2 \, T_j^2 \nonumber \\
  & + & 2 \left( \mathrm{Tr}(\gamma)-2\right) \eta_{\mathrm{APD}} \;T_j +
  4.
  \label{Pjpolynom}
\end{eqnarray}
It becomes clear that the determinant is linked to the second-order
dependence of $P_j^{-2}$ in the transmittance, while
the trace can be directly obtained from the linear dependence of $P_j^{-2}$.
The basic difficulty to estimate $\det(\gamma)$ results from the fact
that the relevant information is hidden in terms of
order $(\eta_{\mathrm{APD}}T)^2$, which are very small
for our experimental data given the low values of $\eta_{\mathrm{APD}}$.
\par

One could then try to increase the overall APD detection
efficiency $\eta_{\mathrm{APD}}$ by releasing either the spatial
or spectral filtering conditions. However, from an experimental
point of view, this does not seem realistic for several reasons.

\par
First, we would move to a region where the physics become
multimode, which is clearly outside the framework of the developed
model. In principle, the photon counting method allows one to
check whether the single-mode description of the experiment is
appropriate or not. If only a single mode is detected, then
$P^{-2}$ should be a quadratic polynomial in
$\eta_{\mathrm{APD}}T$, cfr. Eq.~(\ref{Pjpolynom}). More
generally, if the detector effectively registers light from $N$
modes in a Gaussian state, then $P^{-2}$ becomes a polynomial of
$2N$-th order in $\eta_{\mathrm{APD}} T$ \cite{JFNJCquantph}. So,
after measuring $P$ as a function of $\eta_{\mathrm{APD}} T$ one
could perform a fitting to determine the minimal number of modes
$N$ that is necessary for the description of the observed signal.
However, a successful application of this technique would require
a very high precision in the measurement of $P$ and a high
$\eta_{\mathrm{APD}}$.

A second problem with removing the spatial and/or spectral filters
is that we would loose any possibility to cross-check our results
with classical parametric gain or homodyne measurements. Last,
even in the case of no spatial filter and 10~nm spectral filter,
the overall APD detection efficiency will remain low given our
experimental setup, and we do not expect to gain much according to
our numerical simulations of the constraints on the global
efficiency presented below.

\section{Numerical simulations}   \label{S4}

We have seen that the low APD detection efficiency
$\eta_{\mathrm{APD}}$ precludes a reliable estimate of
$\det(\gamma)$ via the photon-counting method. It is thus
important to determine the efficiency $\eta_{\mathrm{APD}}$ that
should be attained in order to be able to estimate $\det(\gamma)$
with acceptably small errors. More generally, it is interesting to
investigate the dependence of the estimation errors on
$\eta_{\mathrm{APD}}$. For this purpose, we have carried out
extensive numerical simulations of the experiment for several
values of $\eta_{\mathrm{APD}}$, the other parameters of the
simulation being chosen in accordance with the experimental
values. In particular, we have assumed a measurement repetition
rate $N_{\mathrm{rep}}=780.4$~kHz and a total measurement time
$t=100$~s for each transmittance $T_j$. The total number of
measurements for each $T_j$ is then given by
$N_{\mathrm{\mathrm{tot}}}=N_{\mathrm{rep}}t$. We have further
assumed that measurements were carried out for four different
transmittances $T_1=1$, $T_2=0.75$, $T_3=0.5$ and $T_4=0.25$, and
we used the experimentally obtained values $\det(\gamma)=1.156$
and $\mathrm{Tr}(\gamma)=2.321$ as a typical example
(corresponding to a pump average power of 1.21mW).

The determinant and the trace of $\gamma$ were estimated from the
simulated experimental data with the help of the
maximum-likelihood technique described in the preceding section.
Since the ML estimator is generally biased, we define the
deviation of the estimate from the true value as
\begin{eqnarray}
\sigma^2_{\mathrm{det}}&=&\langle(\det(\gamma)_{\mathrm{est}}-\det(\gamma)_{\mathrm{true}})^2\rangle,
\nonumber \\
\sigma^2_{\mathrm{Tr}}&=&\langle(\mathrm{Tr}(\gamma)_{\mathrm{est}}-\mathrm{Tr}(\gamma)_{\mathrm{true}})^2\rangle,
\label{sigma}
\end{eqnarray}
where $\langle\rangle$ indicates averaging over an ensemble of experiments. In
practice, we simulated 1000 times the whole experiment,
from data acquisition to the ML estimation,
and we then calculated (\ref{sigma}) by averaging over the ensemble.
Since the total number of measurements $N_{\mathrm{\mathrm{tot}}}$
was very large, we approximated the binomial
distribution of $C_j$ by a normal distribution with the same mean and
variance.
\par

\begin{figure}[!t]
\resizebox{0.8\hsize}{!}{\includegraphics{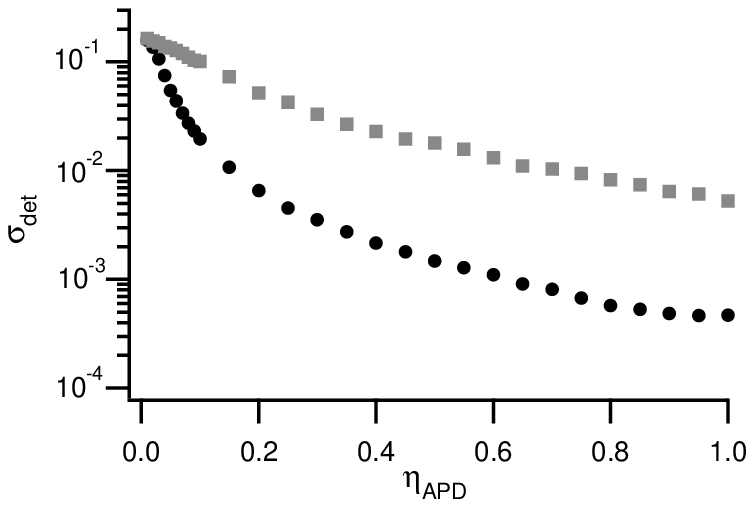}}
\caption{The dependence of the variance $\sigma_{\det}$ of the
estimation of $\det(\gamma)$ on the detector efficiency
$\eta_{\mathrm{APD}}$ in case when the parameters $T_j$ and
$\eta_{\mathrm{APD}}$ are known exactly (circles) and when the
experimental uncertainties of $T_j$ is $0.5$\% and the relative
uncertainty of $\eta_{\mathrm{APD}}$ is $1$\% (squares). See text
for further details. } \label{directfig4}
\end{figure}

Besides the statistical fluctuations of $C_j$ and the intrinsic
difficulty of estimating $\det(\gamma)$ at low detection
efficiencies, other factors contribute to the estimation errors,
namely the uncertainty in the knowledge of $T_j$ and
$\eta_{\mathrm{APD}}$. To isolate the errors stemming from low
$\eta_{\mathrm{APD}}$, we have first assumed that all parameters
$T_j$ and $\eta_{\mathrm{APD}}$ are known precisely, hence the
statistical fluctuations of $C_j$ are the only source of errors.
The resulting $\sigma_{\mathrm{det}}$ is plotted as circles in
Fig.~\ref{directfig4}. For very low $\eta_{\mathrm{APD}}$, the
estimates of $\det(\gamma)$ are randomly distributed in the
interval $[1,(\mathrm{Tr}(\gamma))^2/4]$ and
$\sigma_{\mathrm{det}}\approx
\frac{1}{2}(\mathrm{Tr}^2(\gamma)/4-1)$. The estimation error
rapidly decreases as $\eta_{\mathrm{APD}}$ grows, and our
numerical simulations reveal that a reliable estimate of
$\det(\gamma)$ with $\sigma_{\mathrm{det}}<10^{-2}$ could be
obtained for $\eta_{\mathrm{APD}}>15$\%.

\par
The uncertainties of $T_j$ and $\eta_{\mathrm{APD}}$ significantly
increase the estimation error for higher $\eta_{\mathrm{APD}}$. We
have performed numerical simulations taking into account that
$T_j$'s are known with an uncertainty of $0.5$\%, and the relative
uncertainty of $\eta_{\mathrm{APD}}$ is $1$\% which corresponds to
the actual experimental situation. The resulting
$\sigma_{\mathrm{det}}$ is plotted as squares in
Fig.~\ref{directfig4}. We observe that $\sigma_{\mathrm{det}}$ is
much higher than in the previous case, except for the region of
very small $\eta_{\mathrm{APD}}$. To obtain a satisfactorily
accurate estimate of $\det(\gamma)$ with $\sigma\approx 2\times
10^{-2}$, we need $\eta_{\mathrm{APD}}\gtrsim 50$\%.

\begin{figure}[!ht]
 \resizebox{0.8\hsize}{!}{\includegraphics{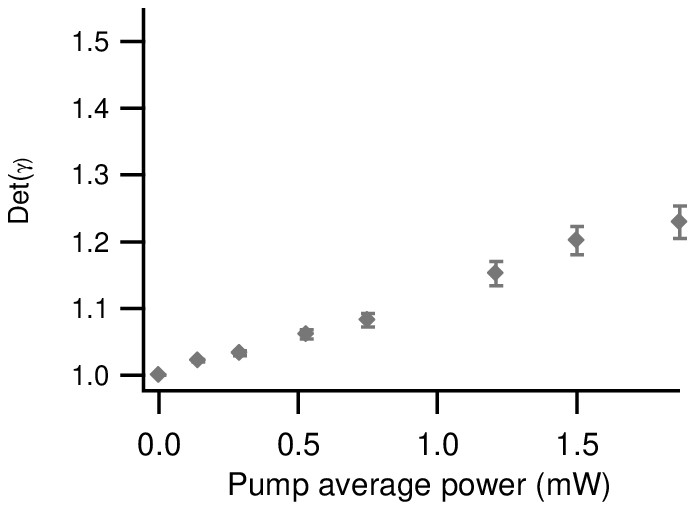}}
\caption{Results of the estimation of $\det(\gamma)$ from
simulated photon counting measurement assuming overall detection
efficiency $\eta_{\mathrm{APD}}=50$\%. This figure has been scaled
so as to be easily compared to the experimental results presented
on Fig.\ref{expresult2}(b).} \label{directfig5}
\end{figure}

In order to demonstrate that $\eta_{\mathrm{APD}}=50$\% is indeed
sufficient for the whole range of values of the pump power, we have
simulated the results of an experiment at
$\eta_{\mathrm{APD}}=50$\% for the same values of the pump power
as in Figs.~\ref{expresult1} and \ref{expresult2}. The results are
given in Fig.~\ref{directfig5} which shows the mean estimated
values of $\det(\gamma)$ as well as the resulting error bars.
We find that these estimates are in very good agreement with the
true values used in the simulation.

Finally, note that our numerical simulations also confirm that the
estimate of $\mathrm{Tr}(\gamma)$ is very accurate: we have found
that $\sigma_{\mathrm{Tr}}\leq 10^{-2}$ even for
$\eta_{\mathrm{APD}}$ as low as 1\%.

\section{Conclusions}

In this paper, we have shown that direct photon-counting
detection can be used, instead of homodyning, to evaluate the squeezing
and purity of an arbitrary single-mode gaussian state.
For the rather generic states that we considered, the trace of the
covariance matrix can be accurately determined, even with an
overall detection efficiency $\eta$  in the percent range, while
its determinant (related to the state purity) requires a much
higher $\eta$, typically around 50\%.

In principle, such efficiencies are well within the reach of
silicon photon-counting avalanche photodiodes, but an important
problem remains : most sources do not emit single-mode gaussian
light, but rather multimode light. This is not a problem when a
homodyne detection is used, because the local oscillator acts
as a very efficient single-mode filter. On the other hand,
a photon counter detects photons in any mode. Therefore, detecting
a good approximation of a single mode state requires appropriate
spatial and spectral filters, respectively obtained from pinholes
and diffraction gratings. Unless a special effort is made, these
filters will have a low overall transmission (a few percent in our
experiment), and thus the direct detection method will fail to
determine the state purity.

In principle, there are various ways for improvement, which are
open for further experimental work. Ideally, the source itself
should emit single mode light, which might be obtained by
appropriate phase-matching conditions in a $\chi^{(2)}$ non-linear
crystal. On the filtering side, interferometric multidielectric
filters provide transmission values which are much higher than
standard slits and grating set-ups. A combination of these various
techniques will be probably needed to reach the high overall
efficiencies needed for many potential applications.

As a conclusion, it appears that a broad  variety of techniques is
available to characterize quantum continuous variables, and that
these methods will certainly continue to develop for applications
in quantum cryptography, quantum communications, and possibly
quantum computing. Perhaps the  most appealing application of the
photon-counting method is the direct determination of the
entanglement of two-mode Gaussian states by measuring only the
purity of the two-mode state and the marginal purities of the
single-mode states on each side \cite{JFNJCquantph,Adesso04}. All
these purities can be determined with the photon counting method
using only local measurements. The distinct feature  of this
approach is that no interferometric stability is required if one
is dealing with squeezed vacuum states, which is the case in many
experiments. This may be an important advantage in the
characterization of entanglement distribution over long-distance
continuous-variable quantum communication networks.

\begin{acknowledgments}
This work is supported by the European IST/FET program. NJC and JF
acknowledge financial support from the Communaut\'e Fran\c{c}aise
de Belgique under grant ARC 00/05-251, from the IUAP programme of
the Belgian government under grant V-18, from the EU under
projects RESQ (IST-2001-37559) and CHIC (IST-2001-33578). JF also
acknowledges support from the  grant LN00A015  of the Czech
Ministry of Education.

\end{acknowledgments}


\end{document}